\documentclass[epj]{svjour}
\usepackage{graphics}
\begin{document}
\title{Application of thermodynamics to driven systems}
\author{R. Mahnke\inst{1} 
\thanks{e-mail: \texttt{reinhard.mahnke@uni-rostock.de}}
\and J. Kaupu\v{z}s\inst{2} \and J. Hinkel\inst{1} \and H. Weber\inst{3}
%
}                     
\offprints{R. Mahnke}          
\institute{
Institute of Physics, Rostock University, D--18051 Rostock, Germany 
\and Institute of Mathematics and Computer Science, 
University of Latvia, LV--1459 Riga, Latvia 
\and Lule\aa{} University of Technology, Department of Physics, 
SE--97187 Lule\aa, Sweden
}
\date{Received: date / Revised version: date}
%
\abstract{
Application of thermodynamics to driven systems is discussed.
As particular examples, simple traffic flow models are considered. On a
microscopic level, traffic flow is described by Bando's optimal velocity
model in terms of accelerating and decelerating forces. It allows to
introduce kinetic, potential, as well as total energy, which is the
internal energy of the car system in view of  thermodynamics. The latter
is not conserved, although it has certain value in any of two possible
stationary states corresponding either to fixed point or to limit cycle
in the space of headways and velocities. On a mesoscopic level
of description, the size $n$ of car cluster is considered as a
stochastic variable in master equation. Here $n=0$ corresponds to the
fixed--point solution of the microscopic model, whereas the limit cycle
is represented by coexistence  of a car cluster with $n>0$ and free flow
phase. The detailed balance holds in a stationary state just like in 
equilibrium liquid--gas system.  It allows to define free energy of the
car system and chemical potentials of the coexisting phases,  as well as
a relaxation to a local or global free energy minimum. In this sense the
behaviour of traffic flow can be described by equilibrium
thermodynamics. We find, however, that the chemical potential of the
cluster phase of traffic flow  depends on an outer parameter -- the
density of cars in the free--flow phase. It allows to distinguish
between the traffic flow as a driven system and purely equilibrium
systems. 
} 
\maketitle
\section{Introduction}
\label{intro}

An extension of thermodynamic concepts from equilibrium to nonequilibrium
or driven systems is one of the fundamental problems in physics. It refers also
to so--called nonphysical systems, like traffic or granular flow, economics, biological
systems, etc., where the laws of microscopic interaction and motion differ from those
known in physics. Different approaches have been developed till now. In the 
geometrical formulation of thermodynamics~\cite{Grmela93}, the latter is regarded
as a theory arising in the analysis of dynamics. In this concept the equilibrium
thermodynamics is represented by a manifold of time--independent equilibrium
states, whereas the thermodynamics of driven system is represented by a manifold
of slowly evolving states. A $k$--component system undergoing chemical reaction
is considered as an example in~\cite{Grmela93}. A more widely discussed approach
is based on the introduction of entropy~\cite{RHM86,Smith05} and usage of the
entropy maximization principle in various applications, e.~g., linear dissipative
driven systems~\cite{Smith05} and single--lane traffic~\cite{RHM86}. An appropriate
definition of temperature is a relevant question when we speak about a
nonphysical system. In~\cite{RHM86} the temperature $T$ and pressure $p$ of traffic flow 
have been introduced via derivatives of certain thermodynamic functions, and it 
has been found that $T$ is negative at typical velocities.
In another approach~\cite{LRM02} similarities between traffic and granular flow
have been discussed proposing two effective temperatures: one characterising
fast or single--car dynamics, and another --- slow or collective dynamics of
traffic flow.

As mentioned in~\cite{RHM86}, entropy need not occupy a position of primacy 
in a general theory beyond the classical equilibrium thermodynamics.  We have found
that in cases where the stationary state of a driven system has the property of
detailed balance in a space of suitable stochastic variable, the thermodynamic potential
can be easily introduced based on this property in a complete analogy with equilibrium
systems. This approach can prove to be useful in many applications due to its
relative simplicity. As an example we consider formation of a car cluster in 
one--lane traffic and show its analogy with the phase separation in 
supersaturated vapour--liquid system. 

The aggregation of particles out of an initially homogeneous situation
is well known in physics, as well as in other branches of natural
sciences and engineering. The formation of bound states as an
aggregation process is related to self--organization
phenomena~\cite{vch99,schw97,usms88}. The formation of car clusters
(jams) at overcritical densities in traffic flow is an analogous
phenomenon in the sense that cars can be considered as interacting
particles~\cite{prhe71,helb04,helb06}. The development of traffic jams in vehicular
flow is an everyday example of the occurrence of nucleation and
aggregation in a system of many point--like cars.  For previous
work focusing on the description of jam formation as a nucleation
process, see~\cite{makau99,makaufr03,physrep,kue02}.
It is related to phase separation and metastability in 
low--dimensional driven systems, a topic which  has attracted much recent 
interest~\cite{Mukamel00,Schutz03,S1,Evans05,KMH05}. Metastability and
hysteresis effects have been observed in real traffic,  see,
e.~g.,~\cite{TGF98,TGF96,cssrev,he01,kerner04,lumawa02,wagner2006}
for discussion of empirical data and the various different 
modelling approaches.

Here we focus on the application of thermodynamics to such a
many--particle system as traffic flow. In a first step we do not 
consider real traffic with its very complicated behaviour but limit our 
investigations to simple models of a directional one--lane vehicular flow. 
We hope this will trigger further development to describe more 
realistic situations of multi--lane traffic as well as of 
synchronized flow~\cite{kerner04}. We have found certain analogy
with physical systems like supersaturated vapour--liquid, although
there are also essential differences, since the traffic flow is a driven
system. We would like to outline some basic ideas and concepts developed
throughout the paper.
\begin{enumerate}
\item
On a microscopic level traffic flow can be described by Bando's optimal
velocity model. In this case the equations of motion can be written as
Newton's law with accelerating and decelerating forces and one can
define the potential $V$ and the kinetic $T$ energy of the car system, 
as well as the total energy $E=T+V$. The latter one has a thermodynamic 
interpretation as $\langle E \rangle = U$, where $U$ is the internal
energy of the system.
\item
Traffic flow is a dissipative system of driven or active particles. It
means that the total energy is not conserved, but we have an energy 
balance equation
\[
\frac{dE}{dt} + \Phi = 0
\]
with the energy flux $\Phi$ following from the equations of motion
and consisting of dissipation (due to friction) and energy input
(due to burning of petrol).
\item
In the long--time limit the many--car system tends to certain stationary
state. In the microscopic description it is either the fixed--point or
the limit cycle in the phase space of velocities and headways depending
on the overall car density and control parameters. The stationary state
is  characterised by certain internal energy.
\item
On a mesoscopic level traffic flow can be described by stochastic master
equation, where stochastic variable is the number of congested cars $n$,
i.~e., the size of car cluster. In this case the fixed--point solution
corresponds to $n=0$, and the limit cycle --- to coexistence of a car
cluster with $n>0$ and free flow phase.
\item
In the space of cluster size, the detailed balance holds for the
stationary solution just like in equilibrium physical systems. It allows
to describe various properties of the stationary state by equilibrium
thermodynamics. In particular, we  calculate free energy of the system
and chemical potentials of coexisting phases in a complete analogy with
the known treatment for a supersaturated liquid--gas system.
\item
In distinction to equilibrium systems, the chemical potential of the
cluster phase of traffic flow is not an internal property of this phase,
since it depends on an outer parameter -- the density of cars in the
free--flow phase. It allows to distinguish between the traffic flow as
a driven system and purely equilibrium systems.
\end{enumerate}

\section{Microscopic optimal velocity model of traffic flow}
\label{sec:BandoII}

Traffic flow can be viewed as a random dynamical
system~\cite{Arnold98,schi02} of active or intelligent
particles~\cite{erdeb00,EbSo05,schw03}. To describe it on a microscopic
level, here we use a simple version of Bando's optimal velocity (OV)
model for point--like cars moving on a one--lane road with periodic
boundary conditions.  The model is defined by the following set of 
equations~\cite{betal94,betal95a,betal95b}
\begin{eqnarray} \label{eq:1}
\frac{dv_i}{dt} &=& \frac{1}{\tau} \left( v_{opt}(\Delta x_i) - v_i \right) \;, \\
\frac{dx_i}{dt} &=& v_i \;,
\end{eqnarray}
where the coordinate $x_i(t)$ as well as the velocity $v_i(t)$ of each
car  $i=1, \dots, N$ at every time moment $t$ can be calculated out of
the initial values by  integrating the coupled equations of motion. Here
\begin{equation} 
v_{opt}(\Delta x) = v_{max} \frac{(\Delta x)^2}{D^2 + (\Delta x)^2} 
\label{eq:vopt}
\end{equation}
is the optimal velocity function depending on the headway distance
$\Delta x_i = x_{i+1} - x_i$ proposed
in~\cite{makau99,makaufr03,physrep,mapi97,makau01}. It includes the 
maximal velocity $v_{max}$ and the interaction distance $D$ as
parameters.  Eq.~(\ref{eq:1}) can be written as
\begin{equation}
m \, \frac{dv_i}{dt} = F_{acc}(v_i) + F_{dec}(\Delta x_i) \;,
\label{eq:acc}
\end{equation}
where 
\begin{eqnarray}
F_{acc}(v_i) &=& 
\frac{m}{\tau} \left( v_{max} - v_i \right) \ge 0 \label{eq:Facc} \\
F_{dec}(\Delta x_i) &=& 
\frac{m}{\tau} \left( v_{opt}(\Delta x_i) - v_{max} \right) \le 0
\label{eq:Fdece}
\end{eqnarray}
are the accelerating and decelerating forces, respectively. 
Similar representation has been introduced already in~\cite{helb04,helb06}.
The only distinguishing feature is that in our case the deceleration
force is specified by~(\ref{eq:Fdece}), whereas in~\cite{helb04}
it is related to a power--like interaction potential. The
coordinate--dependent force term is due to interaction between cars
\begin{equation}
F_{dec}(\Delta x) = v_{max} \frac{m}{\tau} 
\left( \frac{(\Delta x)^2}{D^2 + (\Delta x)^2} - 1 \right)
\end{equation}
and is always negative, starting at $F_{dec}(\Delta x = 0) \; = - v_{max}
m / \tau$, approaching zero at infinite distances. The potential energy
of car system can be defined as $V= \sum_{i=1}^N \phi (\Delta x_i)$,
where $\phi(\Delta x_i)$ is the interaction potential  of the $i$-th car
with the car ahead, which is given by
\begin{equation} 
F_{dec}(\Delta x_i) = - \frac{\partial \phi(x_{i+1} - x_i)}{\partial x_i} 
= \frac{d \phi(\Delta x_i)}{d \Delta x_i}
\label{eq:Fdec}
\end{equation}
By integrating this equation we get 
\begin{equation}
\phi(\Delta x) =  v_{max} \frac{D \, m}{\tau} 
\left[ \frac{\pi}{2} - \arctan\left( \frac{\Delta x}{D} \right) \right] \;,
\end{equation}
where the integration constant is chosen such that 
\linebreak $\phi(\infty)=0$. For comparison, the interaction potential
of the form $\phi(\Delta x) \propto (\Delta x)^{-\alpha}$ has been considered 
in~\cite{helb04}.
Note that $F_{dec}(\Delta x_i)$ is not given by 
$-\partial V/\partial x_i$, since the latter quantity includes an additional
term $- \partial \phi(x_i-x_{i-1}) /\partial x_i$. This term is absent 
in our definition of the force because the car behind does not influence
the motion of the actual $i$--th vehicle. It reflects the fact that,
unlike in physical systems, the third Newton's law does not hold here.

\begin{figure}
\begin{minipage}{10cm}
\resizebox{0.85\textwidth}{!}{%
\includegraphics{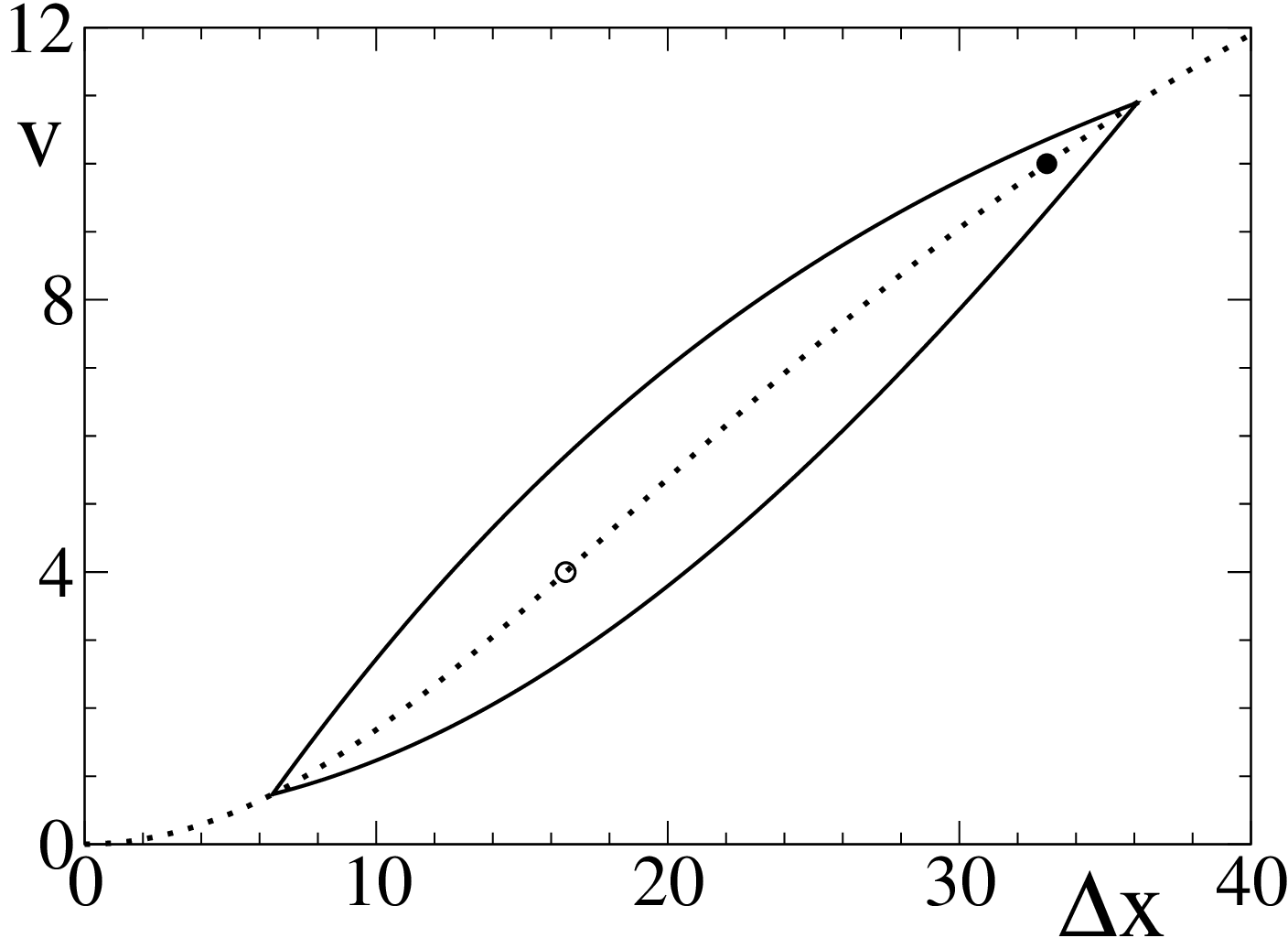}
}
\end{minipage}
\caption{Fixed points (circles) and limit cycle (solid line) in the
space  of headways $\Delta x$ and velocities $v$ of cars. The
solid circle represents the stable fixed point at the car density
$\rho=N/L=0.0303~\mbox{m}^{-1}$. The empty circle is the unstable fixed
point at a larger density $\rho=0.0606~\mbox{m}^{-1}$, where the
long--time trajectory for any car  is the limit cycle shown. The fixed
points lie on the optimal velocity curve (dotted line) given
by~(\ref{eq:vopt}). The parameters are chosen as $N=60$,
$D=33~\mbox{m}$, $v_{max}=20~\mbox{m/s}$, $\tau=1.5~\mbox{s}$, 
and $m=1000~\mbox{kg}$.}
\label{fig:v_dx}
\end{figure}

The total time derivative of the potential energy is
\begin{eqnarray}
\frac{dV}{dt} &=& \sum\limits_{i=1}^N \left[ 
\frac{\partial \phi(\Delta x_i)}{\partial x_i} \frac{dx_i}{dt}
+ \frac{\partial \phi(\Delta x_i)}{\partial x_{i+1}} \frac{dx_{i+1}}{dt} \right] 
\nonumber \\
&=& \sum\limits_{i=1}^N (v_{i+1}- v_i) F_{dec}(\Delta x_i) 
\end{eqnarray}
The total time derivative of the kinetic energy\linebreak $T = \sum_{i=1}^N m
v_i^2/2$ is obtained by multiplying both sides of~(\ref{eq:acc}) by
$v_i$ and summing over $i$. It leads to the following energy balance
equation
\begin{equation}
\frac{dE}{dt} + \Phi = 0
\label{eq:diss}
\end{equation}
for the total energy $E=T+V$ of the car system, where
\begin{equation}
\Phi = - \sum\limits_{i=1}^N \left[ v_i F_{acc}(v_i) +
v_{i+1} F_{dec}(\Delta x_i) \right]
\end{equation}
is the energy flux. It includes both energy dissipation due to friction
and energy input from the engine. Eq.~(\ref{eq:diss}) shows that, in
distinction to closed mechanical systems, the total energy is not
conserved in traffic flow. Nevertheless, it approaches a constant value
in  the long--time limit, where the system converges to one of two
possible stationary states: either to the fixed point $\Delta x_i =
\Delta x_{hom}$, $v_i=v_{opt} \left( \Delta x_{hom} \right)$ (where
$\Delta x_{hom}=L/N$ is the distance between homogeneously distributed
$N$ cars over the road of length $L$), or to the limit cycle in the
phase space of headways and velocities. Both situations are illustrated
in Fig.~\ref{fig:v_dx}. At small enough density of cars there is a
stable fixed point (solid circle), which lies on the optimal velocity
curve (dotted line). An unstable fixed point (empty circle) exists at
larger densities. In the latter case any small perturbation of the
initially homogeneous fixed point situation leads to the limit cycle
(solid line) in the long--time limit.
\begin{figure}
\begin{minipage}{10cm}
\resizebox{0.85\textwidth}{!}{%
\includegraphics{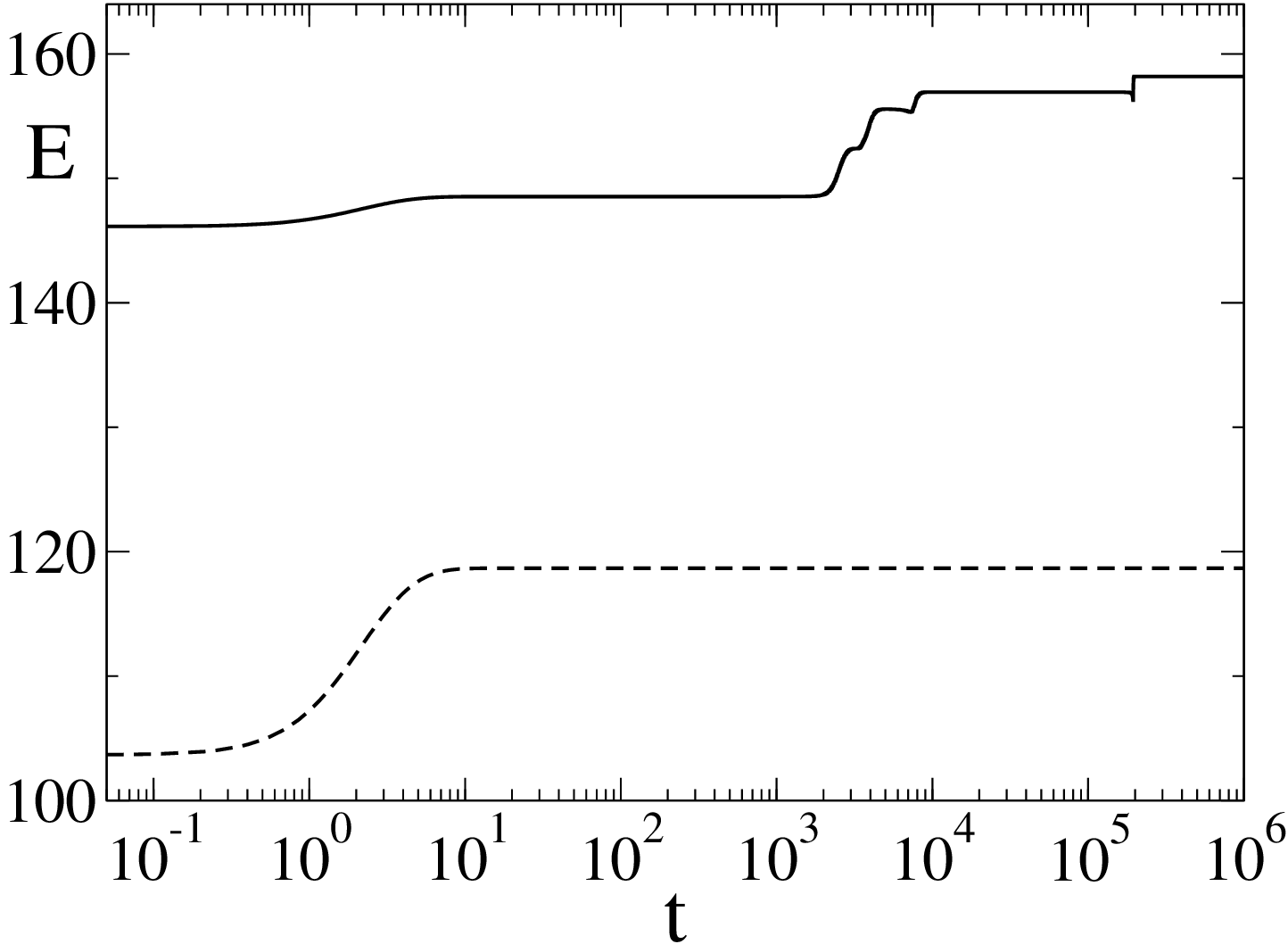}
}
\end{minipage}
\caption{The total energy $E$ of the car system, measured in units of 
$m v_{max}^2/2$, depending on time $t$ given in seconds. The same sets
of parameters have been used as in Fig.~\ref{fig:v_dx}. The upper solid
line corresponds to a larger density $\rho=0.0606~\mbox{m}^{-1}$ where
the limit cycle forms,  whereas the lower dashed line --- to a smaller
density $\rho=0.0303~\mbox{m}^{-1}$ where the convergence to stable
fixed point is observed.}
\label{fig:energy}
\end{figure}

In the thermodynamic interpretation the mean energy $\langle E \rangle$
is the internal energy $U$ of the system. The latter one thus has
certain value in any one of the stationary states. The temporal
behaviour of $E$ for the same sets of parameters as in
Fig.~\ref{fig:v_dx} is shown in Fig.~\ref{fig:energy}. In the case of
the convergence to the limit cycle (solid line) for
$\rho=0.0606~\mbox{m}^{-1}$,  one can distinguish $6$ plateau in the
energy curve. The first one represents the short--time behaviour when
starting from an almost homogeneous initial condition with zero
velocities, and the second plateau is the unstable fixed point
situation. Further on, $4$ car clusters have been formed in the actual
simulation, and this temporal situation is represented by the third
relatively small plateau. The next three plateau with $3$, $2$, and
finally $1$ car clusters reflect the coarse graining or Ostwald ripening
process. The dashed line shows the convergence to the stable fixed point
value at $\rho=0.0303~\mbox{m}^{-1}$.

Apart from the internal energy, other thermodynamic functions can be
introduced as well. In the following sections we will calculate the free
energy $F$ of the traffic flow. By using the known relation $F = U -
T^* S$ we can calculate also the entropy $S$ of traffic flow for a
properly defined `temperature` $T^*$.

Up to now we have considered purely deterministic equations of motion.
Randomness can be included, e.~g., by adding a multiplicative noise
term to~(\ref{eq:acc}). It leads to stochastic differential equations
\begin{eqnarray}
m \, dv_i(t) &=& (F_{acc}(v_i) + F_{dec}(\Delta x_i)) dt  + \sigma \, v_i dW_i(t),
\label{one} \\ 
d x_i(t) &=& v_i dt
\label{two}
\end{eqnarray}
considered and solved numerically in~\cite{physrep}. 
Here $\sigma$ is the noise amplitude, and
$dW_i(t)$ is the increment of Wiener process. Similar equation with additive
noise term has been studied in~\cite{helb04,helb06}. An advantage of the 
version with multiplicative noise is that it guarantees the positiveness
of velocities $v_i$. 
In the deterministic model the departure
(leaving a cluster) times are strongly correlated in such a way that,
in the stationary regime,
one car leaves the cluster after each time interval
of a given length $\tau_1$. The arrival (adding to a cluster) times  
also are strongly correlated due to the repulsive forces.
The noise makes these correlations weaker. It allows to apply the formalism
of stochastic Markov processes to describe approximately the fluctuations
of the cluster size, as discussed in the following section.

\section{Mesoscopic stochastic model of traffic flow}
\label{sec:StochasticsIII}

It is easier to study the formation of a car congestion on a mesoscopic
level, as it has been done
in~\cite{makau99,makaufr03,physrep,mapi97,makau01}, where we do not
follow each individual car, but only look for the number of congested
cars $n$, i.~e., the size of car cluster. In this description it is also
very easy to introduce the randomness, by considering $n$ as a
stochastic variable. 
Following~\cite{makau99,makaufr03,physrep,mapi97,makau01}, in the
simplest model only one cluster on a circular road is considered, and
the probability $p(n,t)$ that it contains $n$ cars at time $t$ is given
by one--step master equation
\begin{eqnarray}
\frac{dp(n,t)}{dt} &=& w_+(n-1) \, p(n-1,t) \nonumber \\ 
                   &+& w_-(n+1) \, p(n+1,t) \\
&-& \left[ w_+(n)+w_-(n) \right] \, p(n,t)  \quad : \quad 0<n<N  \;. \nonumber
\end{eqnarray}
For $n=0$ and $n=N$ the equations look different, i.~e., terms with
$p(-1,t)$ and $p(N+1,t)$ are absent. In the simplest version of the
model of point--like cars the transition rates are $w_-(n)=1/\tau$ and
$w_+(n)= v_{opt} \left( \Delta x_{free} \right)/\Delta x_{free}$, where
$\tau$ is a reaction time constant and $\Delta x_{free}(n) = L/(N-n)$ is the
mean headway distance in the free flow phase. In this model no large
stable cluster forms at low densities of cars, whereas a macroscopic
fraction of them are condensed (jammed) into the cluster above certain
critical density. The first situation corresponds to the fixed--point
solution of the Bando model, whereas the second one  --- to the limit
cycle. It is a remarkable fact that the stationary solution
$p^{st}(n) = \lim_{t \to \infty} p(n,t)$ obeys the detailed balance condition
$p^{st}(n) \, w_+(n) = p^{st}(n+1) \, w_-(n+1)$. It allows to describe
some properties of the model by equilibrium thermodynamics in analogy to
the liquid--vapour system, as discussed in the following sections, in
spite of the fact that the traffic flow is a driven, i.~e.,
nonequilibrium system.

\section{Free energy of the liquid--gas system}
\label{sec:liqgas}

The principle of detailed balance is useful to describe the equilibrium 
in a physical system. 
In particular, we analyse the condensation of supersaturated vapour 
to show how the free energy and chemical potentials can be derived
based on this principle with the following idea to apply
the same scheme for traffic flow. 

For simplicity, we consider a situation where only one cluster of
molecules coexists with the vapour phase. The number $n$ of molecules
called monomers binded in the cluster is a stochastic variable, whereas
their total number $N$ in a given volume $V$ is fixed. The
stochastic events of adding or removing one monomer are characterised by
transition rates $w_+(n)$ and $w_-(n)$ depending on the actual cluster
size $n$. Following~\cite{usms88,physrep}, the detailed balance reads
\begin{equation}
\frac{w_+(n-1)}{w_-(n)}
=\exp \left( -\frac{F(n)-F(n-1)}{k_BT} \right) \;,
\label{dbal}
\end{equation}
where $T$ is temperature, $k_B$ the Boltzmann constant, and $F(n)$ is
the free energy of state (including all possible microscopic
distributions of coordinates and momenta of free monomers)  with cluster
size $n$. For large enough $n$ (\ref{dbal}) can be approximated as
\begin{equation}
\frac{w_+(n)}{w_-(n)} 
\simeq \exp \left( -\frac{\partial F/\partial n}{k_BT} \right) \;,
\label{wpm_appr}
\end{equation}
which leads to the equation
\begin{equation}
\ln \left[ \frac{w_+(n)}{w_-(n)} \right] = - \frac{1}{k_BT} 
\, \frac{\partial F}{\partial n} \;.
\label{eq:ln}
\end{equation}
From this we get
\begin{equation}
F=F_0 - k_BT \int\limits_0^n \ln \left[ \frac{w_+(n')}{w_-(n')} \right] dn' \;, 
\label{Fgen}
\end{equation}
where $F_0=F(n=0)$ does not depend on the cluster size $n$. It is the
free  energy of the system without cluster, in this case the free energy
of an ideal gas. We insert here the physical ansatz for the transition
rates (see~\cite{physrep})
\begin{equation}
\frac{w_+(n)}{w_-(n)} = \frac{\lambda_0^3(T)(N-n)}{V}
\exp \left( \frac{f_{n-1}(T)-f_n(T)}{k_BT} \right) \,,
\label{wpm}
\end{equation}
where $V$ is the fixed volume of the system, $f_n(T)$ is the binding energy
of a cluster of size $n$,\linebreak $\lambda_0(T) = h/(2 \pi m k_BT)^{1/2}$ 
is the de Broglie wave length of a monomer, and $h$ is the Planck's constant.
By using the approximation 
$f_{n-1}(T)-f_n(T) \simeq - \partial f_n(T) / \partial n\, ,$
we obtain
\begin{equation}
F = F_0 - k_BT \int\limits_0^n
\ln \left[ \frac{\lambda_0^3(T)(N-n')}{V} \right] dn' 
+ f_n(T) \;.
\end{equation}
The integration, using $\int \ln x \, dx = x \ln x -x$, yields
\begin{eqnarray}
\label{F}
F &=& F_0 - k_BTN \left[ \ln \left( \lambda_0^3(T) \frac{N}{V} \right) -1 \right] \\
&+& k_BT (N-n) \left[ \ln \left( \lambda_0^3(T) \frac{N-n}{V} \right) -1 \right]
+ f_n(T) \nonumber \;.
\end{eqnarray}
The free energy of ideal system (gas) $F_0$ cannot be obtained from the
detailed balance relation. It is given by $F_0=-k_BT \ln Z_{id}$,
where $Z_{id}$ is the partition function of the ideal gas
\begin{eqnarray}
Z_{id} &=& \frac{1}{N!} \prod_{\alpha=1}^{3N} \frac{1}{h}
\int\limits_0^L dx_{\alpha} \int\limits_{-\infty}^{\infty} dp_{\alpha}
\, \exp \left( -\frac{p_{\alpha}^2}{2m k_BT} \right) \nonumber \\
&=& \frac{1}{N!} \left( \frac{L}{\lambda_0(T)} \right)^{3N} \;.
\label{eq:Zid}
\end{eqnarray}
Hence, applying the Stirling formula $\ln N! \simeq N \ln N -N$, we obtain
\begin{equation}
F_0 = k_BTN \left[ \ln \left( \lambda_0^3(T) \frac{N}{V} \right) -1 \right] \;.
\label{F0}
\end{equation}
By inserting~(\ref{F0}) into~(\ref{F}) we recover the known expression
\begin{equation}
F = k_BT (N-n) \left[ \ln \left( \lambda_0^3(T) \frac{N-n}{V} \right) -1 \right]
+ f_n(T)
\label{eq:F}
\end{equation} 
for the free energy of liquid--gas system under isothermal and isochoric
conditions. The binding energy $f_n(T)$ can be written as
\begin{equation}
f_n(T) = \mu_{\infty}(T) n + \sigma A(n) \;,
\end{equation}
where $\mu_{\infty}(T) n$ represents the volume contribution,
\linebreak 
$\mu_{\infty}(T)<0$ being the chemical potential for a flat droplet
interface (with infinite radius $r$), and $\sigma A(n)$  is the surface
contribution. Here $\sigma>0$ is the surface tension, whereas $A(n)=4
\pi r^2$ is the surface area of a droplet with radius $r$. Taking into
account that the number of particles (molecules) in the cluster is $n=
\left( c_{\mathrm clust} \, 4 \pi/3 \right) r^3$, where $c_{\mathrm
clust}$ is the particle density inside the cluster, Eq.~(\ref{F}) can be
written as
\begin{eqnarray}
\frac{F-F_0}{V k_BT} &=& \rho \left\{ \left( 1- \frac{n}{N} \right) 
\left[ \ln \left( 1- \frac{n}{N} \right) -1 \right] + 1 \right. \nonumber \\
&-& \left. \frac{n}{N} \ln \left( \lambda_0^3(T) \rho \right) 
 + \frac{\mu_{\infty}(T)}{k_BT} \frac{n}{N} \right. \label{eq:Fdif} \\
&+& \left. \frac{3}{2} \ell(T)
\left( c_{\mathrm clust} \, 4 \pi/3 \right)^{1/3} N^{-1/3} 
\left( \frac{n}{N} \right)^{2/3} \right\} \nonumber \;,
\end{eqnarray}
where  $\rho = N/V$ is the overall density and
\begin{equation}
\ell(T) = \frac{2 \sigma}{c_{\mathrm clust} k_BT} 
\end{equation}
is the diffusion length (width) of the liquid--gas interface.

\begin{figure}
\begin{minipage}{10cm}
\resizebox{0.85\textwidth}{!}{%
\includegraphics{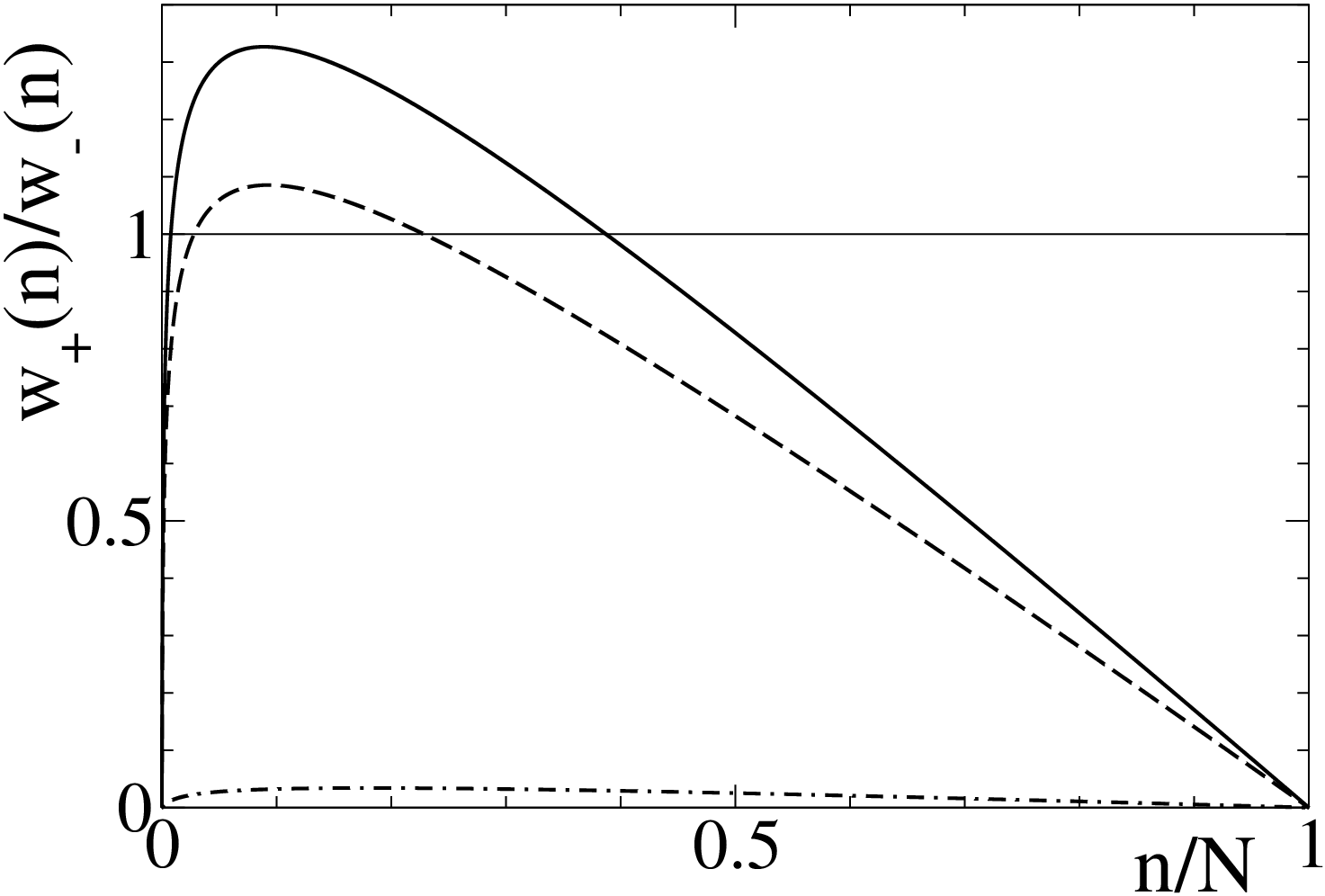}
}
\end{minipage}
\caption{The ratio of transition rates $w_+(n)/w_-(n)$ depending on the
fraction of condensed particles  $n/N$ for three dimensionless densities  
$\tilde \rho=5 \cdot 10^{-7}$ (dot--dashed line), $\tilde \rho=10^{-5}$
(dashed line), and $\tilde \rho=1.2 \cdot 10^{-5}$ (solid line).}
\label{fig:wpm}
\end{figure}

\begin{figure}
\begin{minipage}{10cm}
\resizebox{0.85\textwidth}{!}{%
\includegraphics{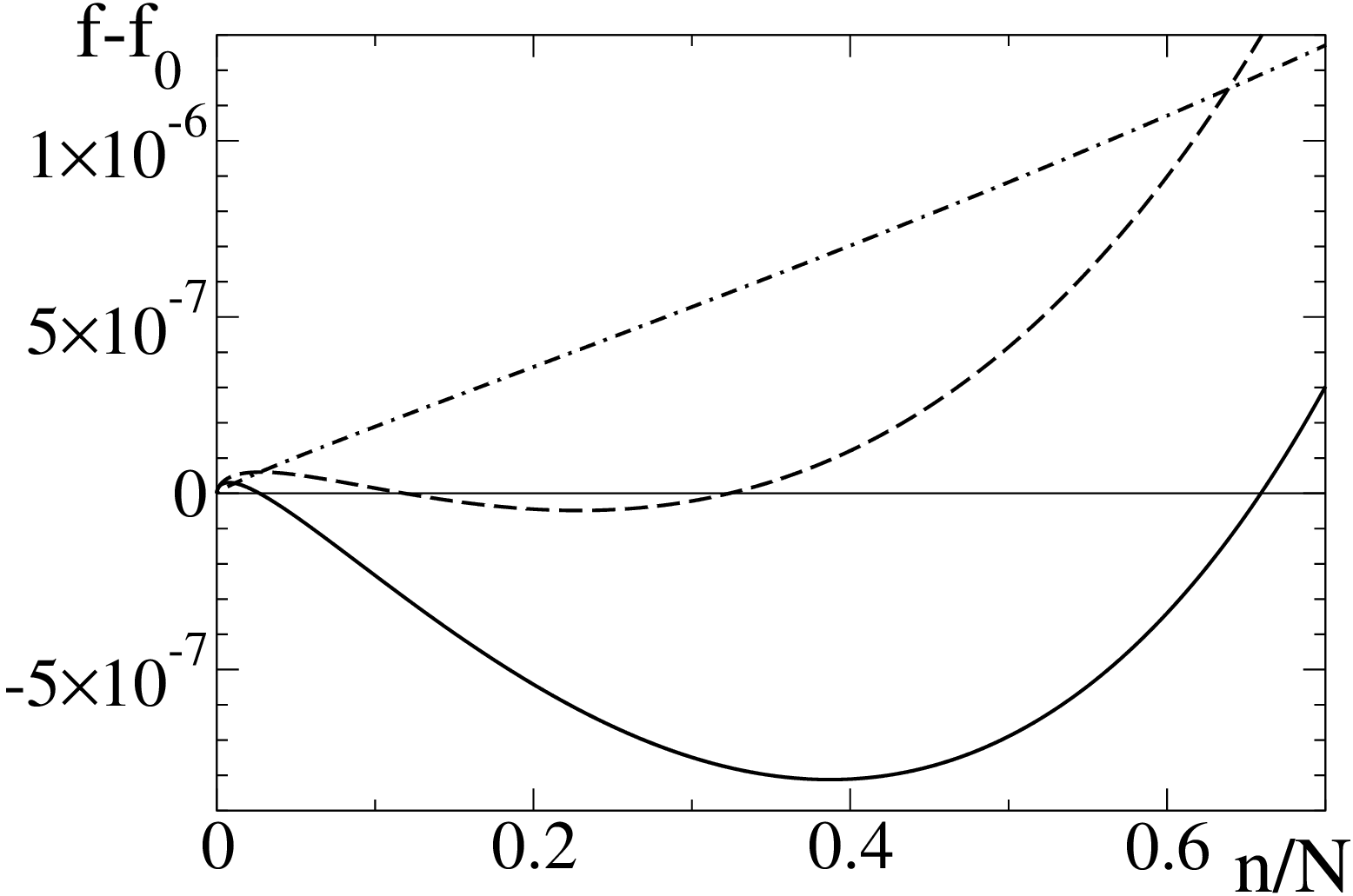}
}
\end{minipage}
\caption{Normalised free energy difference $(F-F_0)/(\widetilde V k_BT)=f-f_0$
depending on the fraction of condensed particles 
$n/N$ for three dimensionless densities  
$\tilde \rho=5 \cdot 10^{-7}$ (dot--dashed line), $\tilde \rho=10^{-5}$ (dashed 
line), and $\tilde \rho=1.2 \cdot 10^{-5}$ (solid line).}
\label{free_en}
\end{figure}

Further on we introduce dimensionless density 
$\tilde \rho = \lambda_0^3(T) \rho$ and dimensionless volume
$\widetilde V = V/\lambda_0^3(T)$. 
In this notation the equation~(\ref{wpm}) transforms to 
\begin{eqnarray}
\label{wpmd}
&&\frac{w_+(n)}{w_-(n)} = \tilde \rho \left( 1- \frac{n}{N} \right)
\, \exp \left( - \frac{\mu_{\infty}(T)}{k_BT} \right) \\
&\times& \! \! \exp \left(  - \ell(T)
\left( c_{\mathrm clust} \, 4 \pi/3 \right)^{1/3} 
\widetilde V^{-1/3} \tilde \rho^{-1/3}
\left( \frac{n}{N} \right)^{-1/3} \right) \nonumber \,,
\end{eqnarray}
whereas~(\ref{eq:Fdif}) becomes
\begin{eqnarray}
&&\frac{F-F_0}{\widetilde V k_BT} = \tilde \rho \left\{ \left( 1- \frac{n}{N} \right) 
\left[ \ln \left( 1- \frac{n}{N} \right) -1 \right] \right. \nonumber \\
&+& 1 - \left. \frac{n}{N} \ln \left( \tilde \rho \right) 
+ \frac{\mu_{\infty}(T)}{k_BT} \frac{n}{N} \right. \\ 
&+& \left. \frac{3}{2} \ell(T)
\left( c_{\mathrm clust} \, 4 \pi/3 \right)^{1/3} 
\widetilde V^{-1/3} \tilde \rho^{-1/3}  
\left( \frac{n}{N} \right)^{2/3} \right\} \nonumber \;.
\label{F_diml}
\end{eqnarray}
These equations allow us to calculate the ratio
\linebreak  $w_+(n)/w_-(n)$, as
well as  the normalised (dimensionless) free  energy difference
$(F-F_0)/(\widetilde V k_BT)$ depending on the fraction of condensed
molecules $n/N$ at a given overall density for fixed volume and
temperature. The results of calculation for three different
dimensionless densities $\tilde \rho=5 \cdot 10^{-7}, 10^{-5}, 1.2 \cdot
10^{-5}$ at the  values of dimensionless control parameters
$\mu_{\infty}/(k_BT)=-12$ and \linebreak  
$\ell(T) \left( c_{\mathrm clust} \, 4
\pi/3 \right)^{1/3} \widetilde V^{-1/3} = 0.003$ are shown in
Figs.~\ref{fig:wpm}  and~\ref{free_en}.  Note that the extrema of
$F-F_0$ in Fig.~\ref{free_en} correspond to the crossing points with the
horizontal line $w_+(n)/w_-(n)=1$ in Fig.~\ref{fig:wpm}. At the smallest
density (dot--dashed line) there are no crossing points and the free
energy is a monotonously increasing function of $n/N$, showing that the
stable state of the liquid--gas system contains no liquid droplet. 
Stable droplet appears at larger densities (dashed and solid lines) by
overcoming a nucleation barrier (local free energy maximum in
Fig.~\ref{free_en}).

The parameters we have chosen are quite realistic, i.~e.,
comparable with those of water at $T=300~\mbox{K}$ and $V=5 \cdot
10^{-23}~\mbox{m}^3$ with about $37\,250$ molecules (mass $m=2.99 \cdot
10^{-23}~\mbox{kg}$) at $\tilde \rho = 10^{-5}$. For water at
$T=300~\mbox{K}$ we have $\lambda_0(T)= 2.377 \cdot 10^{-11}~\mbox{m}$
and $c_{\mathrm clust} = 3.346 \cdot 10^{28}~\mbox{m}^{-3}$. Hence, the
dimensionless density in the cluster  $\tilde \rho_{\mathrm clust} =
c_{\mathrm clust} \lambda_0^3=4.491 \cdot 10^{-4}$ exceeds about $50$
times the critical mean density  $\tilde \rho = \tilde \rho_{c} \simeq
9.2 \cdot 10^{-6}$ at which  the condensation (i.~e., a minimum of free
energy at $n/N >0$) appears in our  calculation. Assuming the above
parameters of water, we obtain $\ell(T)=8.953 \cdot 10^{-10}~\mbox{m}$
for the width of the liquid--gas interface and surface tension $\sigma 
= 6.20 \cdot 10^8~\mbox{Nm/m}^2$. It is about $3$ times the
characteristic intermolecular distance in the cluster, which is roughly 
$c_{\mathrm clust}^{-1/3} \simeq 3.1 \cdot 10^{-10}~\mbox{m}$. The
critical density $\tilde \rho_{c}$ increases with temperature and
becomes equal to the cluster density $\tilde \rho_{\mathrm clust}$ at
the critical temperature $T=T_c$. In our description the physically
meaningful densities are restricted by $\tilde \rho \le \tilde
\rho_{\mathrm clust}$. It means that no condensation  phase transition
takes place for these physical densities at $T>T_c$. Assuming that
$\mu_{\infty}$ and $\sigma$ do not change with temperature and that the
above given values of dimensionless control parameters correspond to
$T=300~\mbox{K}$, we find $T_c \simeq 430~\mbox{K}$ in our example.

\section{Free energy of traffic flow}

Now we make similar calculation of free energy for the traffic flow
model introduced in Sec.~\ref{sec:StochasticsIII}. Similar general
relations~(\ref{dbal}) to~(\ref{Fgen}) are valid to describe the
stationary (quasi--equilibrium) properties in the space of car cluster
size $n$, since the detailed balance is the property of the stationary
solution of the one--dimensional one--step master equation for the
probability distribution over $n$. Here we only replace $k_B T$ with
$T^*$ which is the `tempe\-rature` of traffic flow having energy
dimension.

The ratio of transition rates in this case reads
\begin{equation}
\frac{w_+(n)}{w_-(n)} = \tau \frac{v_{opt}(\Delta x_{free})}{\Delta x_{free}} \;,
\end{equation}
where $v_{opt}(\Delta x)$ is the optimal velocity function given
by~(\ref{eq:vopt}) and $\Delta x_{free}$ is the headway distance in free
flow phase. Assuming the model of point--like cars we have $\Delta
x_{free}(n) = L/(N-n)$, where $L$ is the length of the road and $N$ is the
total number of cars. It yields
\begin{equation}
\frac{w_+(n)}{w_-(n)} = v_{max} \tau \rho \; \frac{1-n/N}{1+ \rho^2 D^2 (1-n/N)^2} \;,
\end{equation}
where $\rho = N/L$ is the car density. Introducing the dimensionless density
$\tilde \rho = \rho D$ and a dimensionless control parameter
$\tilde b = D/(v_{max} \tau)$, it becomes
\begin{equation}
\frac{w_+(n)}{w_-(n)} = \frac{1}{\tilde b} \, \frac{\tilde \rho (1-n/N)}
{1+ \tilde \rho^2 (1-n/N)^2} \;,
\end{equation}
or
\begin{eqnarray}
\ln \left[ \frac{w_+(n)}{w_-(n)} \right] 
&=& \ln \left( \frac{\tilde \rho}{\tilde b} \right)
+ \ln \left[ 1- \frac{n}{N} \right] \nonumber \\
&-& \ln \left[ 1 + \tilde \rho^2 
\left( 1- \frac{n}{N} \right)^2 \right] \;.
\label{eq:lnw}
\end{eqnarray}
By inserting the latter relation into~(\ref{Fgen}) (where $k_BT \to
T^*$), the integration using $\int \ln \left(1+x^2 \right) dx = 2
\arctan x + x \ln \left( 1+x^2 \right) - 2x$ yields
\begin{eqnarray}
&&\frac{F-F_0}{\widetilde L \, T^*} = \tilde \rho \left\{
\left( 1- \frac{n}{N} \right) \ln \left( 1- \frac{n}{N} \right)
- \frac{n}{N} - \frac{n}{N} \ln \left( \frac{\tilde \rho}{\tilde b} \right) \right. 
\nonumber \\
\label{F_traff}
&&- \left. \left( 1- \frac{n}{N} \right) \ln \left( 1+ \tilde \rho^2
\left[ 1- \frac{n}{N} \right]^2 \right) + \ln \left( 1+ \tilde \rho^2 \right) \right\} 
\nonumber \\
&&+ 2 \arctan \tilde \rho - 
2 \arctan \left( \tilde \rho \left[1- \frac{n}{N} \right] \right) \;,
\end{eqnarray}
where $\widetilde L = L/D$ is the dimensionless length of the road.

\begin{figure}
\begin{minipage}{10cm}
\resizebox{0.85\textwidth}{!}{%
\includegraphics{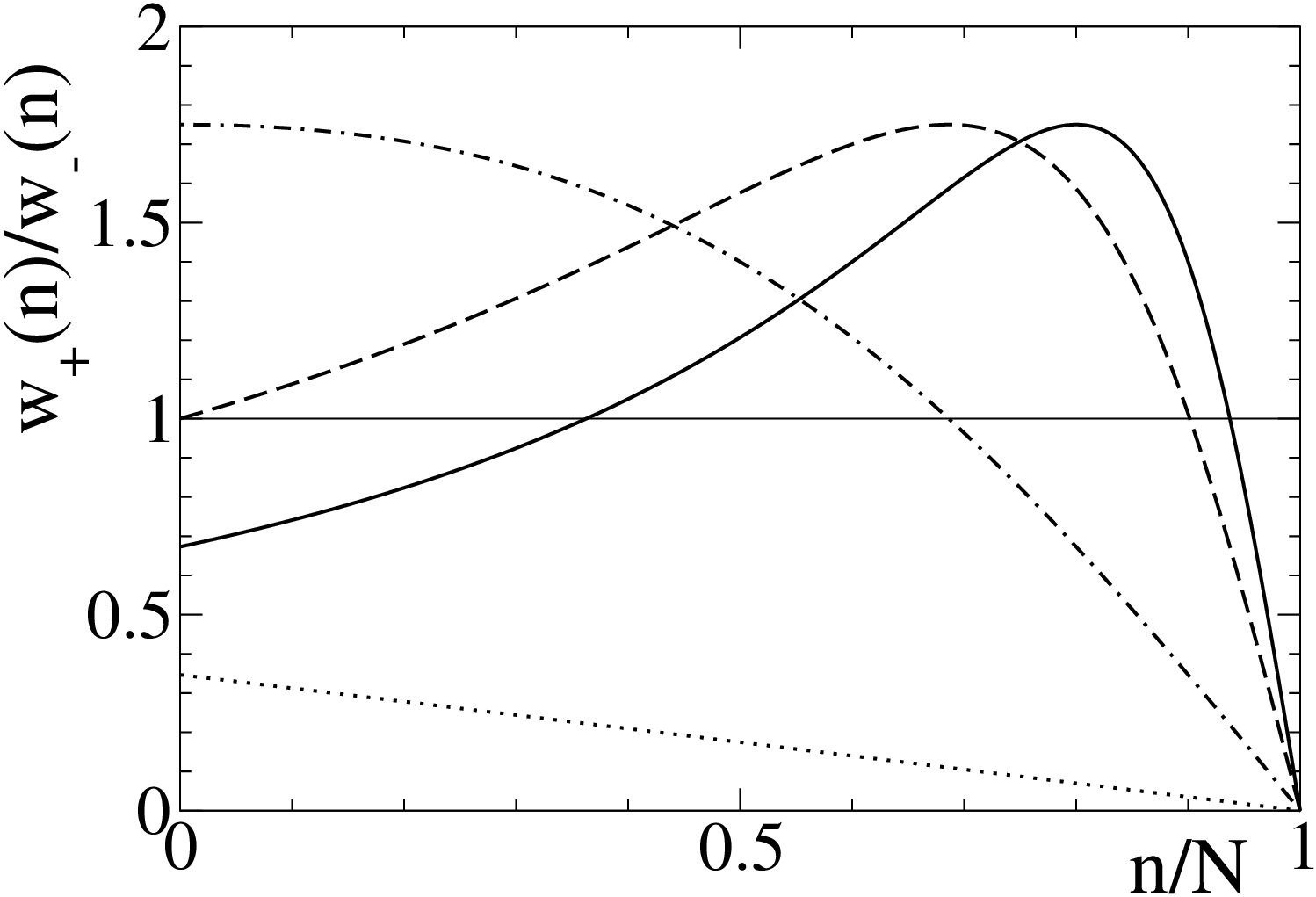}
}
\end{minipage}
\caption{The ratio of transition rates $w_+(n)/w_-(n)$ depending on the
fraction of  congested cars $n/N$ for four dimensionless densities  
$\tilde \rho=0.1$ (dotted line), $\tilde \rho=1$ (dot--dashed  line),
$\tilde \rho=3.186$ (dashed line), and $\widetilde \rho = 5$ (solid line).}
\label{wpm_traff}
\end{figure}

\begin{figure}
\begin{minipage}{10cm}
\resizebox{0.85\textwidth}{!}{%
\includegraphics{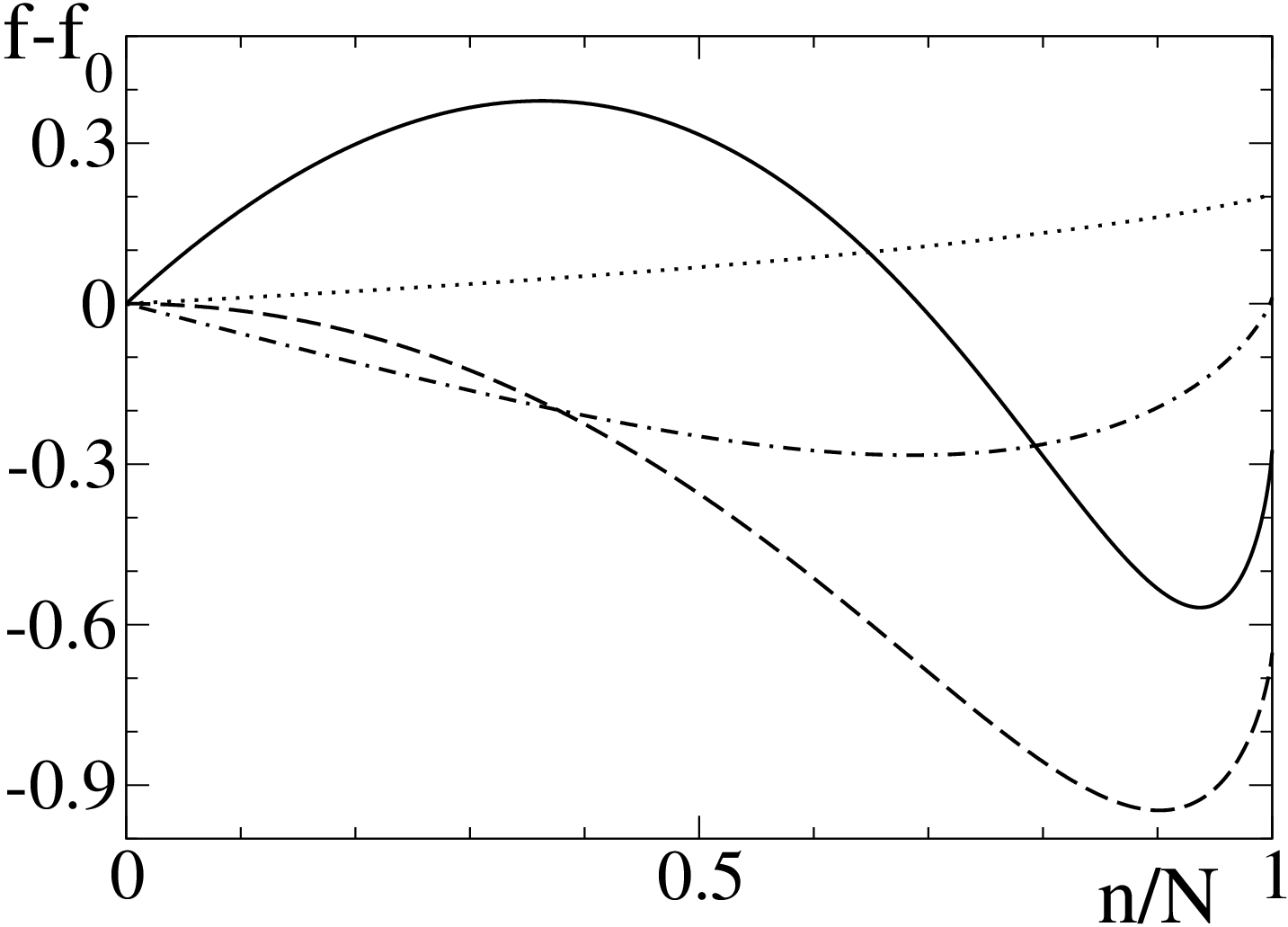}
}
\end{minipage}
\caption{Normalised free energy difference $(F-F_0)/(\widetilde L
T^*)=f-f_0$ depending on the fraction of congested cars  $n/N$ for four
dimensionless densities   $\tilde \rho=0.1$ (dotted line), $\tilde
\rho=1$ (dot--dashed  line), $\tilde \rho=3.186$ (dashed line), and
$\tilde \rho = 5$ (solid line).}
\label{free_en_traff}
\end{figure}

The results for $w_+(n)/w_-(n)$ and $(F-F_0)/(\widetilde L T^*)$
depending on the fraction of congested cars $n/N$ at four different 
densities are shown in Figs.~\ref{wpm_traff} and~\ref{free_en_traff}.
The value of the dimensionless control parameter has been chosen
$\tilde b=2/7 \approx 0.2857$. It corresponds, e.~g., to
$D=24~\mbox{m}$,  $v_{max}=42~\mbox{m/s}$, and $\tau= 2~\mbox{s}$. Like
in the case of the liquid--supersaturated vapour system, 
$w_+(n)/w_-(n)$ is never $1$ and no stable car cluster forms at  small
densities (dotted line). In distinction to the liquid--vapour case, the
cluster appears without a nucleation barrier in the actual traffic flow
model at somewhat larger densities $\tilde \rho$ (dot--dashed line),
whereas the nucleation barrier (free energy maximum) shows up only at
even larger $\tilde \rho$ values (solid line).

In the above calculation we have determined only the difference $F-F_0$,
but not the free energy $F_0$ of the ideal system without car cluster.
Like in the case of supersaturated vapour, the latter cannot be derived
from  the detailed balance. It should be calculated from a microscopic
model. Now, however, we should take into account that the distribution
over momenta for cars is not the same as  that for molecules in an ideal
gas. As a first approximation we may assume similar Gaussian
distribution with only shifted mean value  $\langle p \rangle = m
\langle v \rangle = m v_{opt} \left(\Delta x_{hom} \right)$ in
accordance with the optimal velocity $v_{opt} \left(\Delta x_{hom}
\right)$ in the homogeneous flow of cars with the mean headway distance
$\Delta x_{hom} = L/N = 1/\rho$. The Gaussian form of the distribution
is well consistent with the simulation results for the stochastic
car--following models~\cite{helb04,helb06,physrep}.
We should take into account also that
cars are moving always in one direction, i.~e., momentum $p>0$ always
holds. Finally, in distinction to the ideal gas of molecules, the
coordinates and momenta of cars are one--dimensional. Hence, by analogy
to~(\ref{eq:Zid}) we can write
\begin{eqnarray}
Z_{id} &=& \frac{1}{N!} \prod_{\alpha=1}^{N} \frac{1}{h}
\int\limits_0^L dx_{\alpha} \int\limits_0^{\infty} dp_{\alpha}
\, \exp \left( -\frac{ \left( p_{\alpha} - \langle p \rangle \right)^2}{2m T^*} \right)
\nonumber \\
&\approx& \frac{1}{N!} \left( \frac{L}{\lambda_0(T^*)} \right)^N \;,
\label{Zid_traff}
\end{eqnarray}
where $\lambda_0(T^*) = h/(2 \pi m T^*)^{1/2}$. The latter approximate
equality in~(\ref{Zid_traff}) holds when $\langle p \rangle^2/(2m T^*)
\gg 1$ or, in other words, when the width of the velocity distribution
is narrow as compared to  the mean velocity. The latter condition is well satisfied
for the model~(\ref{one}) --- (\ref{two}) with certain set of control parameters used
in the simulations of~\cite{physrep} (see Fig.~40 there). The distribution width,
however, increases with the noise amplitude. In fact, the 
approximation~(\ref{Zid_traff}) is good enough when the distribution function has
small value at zero momentum $p=0$, as in
the simulation results of~\cite{helb04,helb06}.  According to the above
consideration, temperature in traffic flow is a parameter which controls
this distribution width or the amplitude of velocity and momentum
fluctuations, like in the ideal gas of molecules. According
to~(\ref{Zid_traff}), the ideal part of free energy reads
\begin{eqnarray}
F_0 = -T^* \left[ N \ln \left( L/\lambda_0(T^*) \right) -\ln N! \right]
\nonumber \\
\simeq  T^* N \left[ \ln \left( \rho \lambda_0(T^*) \right) -1 \right] \;.
\label{F0_tr}
\end{eqnarray}
This expression is analogous to~(\ref{F0}).

\section{Relaxation to a free energy minimum}

Now we consider the general behaviour of a system (equally valid for
liquid--gas system with $T^* = k_BT$ and traffic flow) in vicinity of a
local or global minimum of $F(n)$. In this case the argument of exponent 
in~(\ref{wpm_appr}) is small and we can make a Taylor expansion
\begin{equation}
\frac{w_+(n)}{w_-(n)} 
\simeq \exp \left( -\frac{\partial F/\partial n}{T^*} \right)
\simeq 1 - \frac{1}{T^*} \, \frac{\partial F}{\partial n}\;.
\label{wpm_approx}
\end{equation}
It can be rewritten as
\begin{equation}
w_+(n)-w_-(n)
\simeq  - \frac{w_-(n)}{T^*} \, \frac{\partial F}{\partial n} \;.
\label{wp_wm}
\end{equation}
On the other hand, we can write in a deterministic approximation
\begin{equation}
\frac{dn}{dt} = w_+(n) - w_-(n) \;.
\label{kin_eq}
\end{equation}
Comparing~(\ref{wp_wm}) and~(\ref{kin_eq}), we obtain
\begin{equation}
\frac{dn}{dt} 
\simeq  - \frac{w_-(n)}{T^*} \, \frac{\partial F}{\partial n} \;.
\label{eq:kin}
\end{equation}
Like in the Landau theory of phase transitions, we can expand the
free energy around the minimum point $n=n_0$ defined by
\begin{equation}
\left. \frac{\partial F}{\partial n} \right|_{n=n_0} = 0 \;. 
\end{equation}
In the first approximation, where we retain only the leading term,
we have also $w_+(n) = w_-(n) = w_{\pm} \left( n_0 \right)$. It leads
to the kinetic equation
\begin{equation}
\frac{dn}{dt} \simeq - \Gamma_0 \, \left( n -n_0 \right) \;,
\end{equation}
where 
\begin{equation}
\Gamma_0 =  \frac{w_{\pm} \left( n_0 \right)}{T^*} \, 
\left. \frac{\partial^2 F}{\partial n^2} \right|_{n=n_0}
\end{equation}
is the relaxation rate. For $\Gamma_0>0$, what corresponds to
minimum of $F$, the solution is the exponential
relaxation to $n=n_0$, i.~e.,
\begin{equation}
n(t) = n_0 + \left( n(0) -n_0 \right) e^{-\Gamma_0 \, t} \;.
\end{equation}
This solution is valid also for $\Gamma_0<0$, in which case
$n_0$ corresponds to a free energy maximum. In this case it
describes the deviation from this maximum point.

\section{Chemical potentials}

Our system can be considered as consisting of two phases:
the cluster phase with $n$ particles and free energy $F_{cl}(n)$,
and the ideal gas phase with $N_{id}=N-n$ particles and free
energy $F_{id} \left( N_{id} \right)$. The total free energy then 
is $F = F_{cl} + F_{id}$. While the total number of particles $N$
is fixed, the number of particles in any of the phases fluctuates.
According to the definition, we can write 
$\mu_{cl} = \partial F_{cl}/\partial n$ and 
$\mu_{id} = \partial F_{id}/\partial N_{id} 
= - \partial F_{id}/\partial n$ for the chemical potentials of these
phases. Hence 
\begin{equation}
\frac{\partial F}{\partial n} = \frac{\partial F_{cl}}{\partial n}
+ \frac{\partial F_{id}}{\partial n} = \mu_{cl} - \mu_{id}
\label{eq:Fder}
\end{equation}
and the kinetic equation~(\ref{eq:kin}) can be written as
\begin{equation}
\frac{dn}{dt} 
\simeq  - \frac{w_{\pm} \left( n_0 \right) }{T^*} \, 
\left( \mu_{cl} - \mu_{id} \right) \;.
\end{equation}
The latter equation has certain physical interpretation: the
driving force pushing the system to the phase equilibrium is
the difference of chemical potentials in both phases. The
equilibrium is reached when the chemical potentials
of the coexisting phases are equal, i.~e., $\mu_{cl}= \mu_{id}$.

For the liquid--gas system $F_{id}(T,V,N,n)$ is given by (\ref{F0}),
where $N$ is replaced with $N_{id}=N-n$, i.~e.,
\begin{equation}
F_{id}(T,V,N,n) = 
k_BT (N-n) \left[ \ln \left( \lambda_0^3(T) \frac{N-n}{V} \right) -1 \right] \;.
\label{Fideal}
\end{equation}
Hence the total free energy~(\ref{eq:F}) can be written as
\begin{equation}
F = F_{id}(T,V,N,n) + f_n(T) \;.
\end{equation}
The chemical potential of the liquid phase is thus given by the
derivative of the binding energy $f_n(T) \equiv F_{cl}(T,V,N,n)$, i.~e.,
\begin{equation}
\mu_{cl} = \mu_{\infty}(T) + \sigma \, \frac{\partial A(n)}{\partial n}
= \mu_{\infty}(T) + k_BT \, \ell(T) k(n) \;,
\end{equation}
where 
\begin{equation}
k(n) = \frac{1}{r}= 
\left( c_{\mathrm clust} \, 4 \pi /3 \right)^{1/3} n^{-1/3}
\end{equation}
is the curvature of the liquid surface for a droplet with radius $r$
and surface area $A(n)= 4 \pi r^2$.
The chemical potential of the gaseous phase calculated from~(\ref{Fideal}) is
\begin{eqnarray}
\mu_{id} &=& -\frac{\partial F_{id}}{\partial n} = k_BT \, 
\ln \left( \lambda_0^3(T) \frac{N-n}{V} \right) \nonumber \\
&=& k_B \, T \ln \left( \tilde \rho_{gas} \right) \;,
\end{eqnarray}
where $\tilde \rho_{gas} = \lambda_0^3(T)(N-n)/V$ is the dimensionless 
density of molecules in the gaseous phase.
According to these expressions for the chemical potentials,
the ansatz~(\ref{wpm}) can be written as
\begin{eqnarray}
\frac{w_+(n)}{w_-(n)} &=& \exp \left( \frac{\mu_{id} }{k_BT}  \right)
\, \exp \left( \frac{f_{n-1}(T)-f_n(T)}{k_BT} \right) \nonumber \\
&\simeq& \exp \left( - \frac{\mu_{cl}-\mu_{id} }{k_BT}  \right) \;.
\label{eq:wpm}
\end{eqnarray}
The latter relation is consistent with~(\ref{wpm_appr}) and~(\ref{eq:Fder}).

By analogy, the free energy of the free flow phase in traffic is
\begin{equation}
F_{id}(T^*,L,N,n) = 
T^* (N-n) \left[ \ln \left( \lambda_0(T^*) \frac{N-n}{L} \right) -1 \right] \;,
\label{Fideal_tr}
\end{equation}
as consistent with~(\ref{F0_tr}) where we put $N \to N_{id}=N-n$ and
$\rho \to N_{id}/L$.
From~(\ref{Fideal_tr}) we get
\begin{eqnarray}
\mu_{id} &=& -\frac{\partial F_{id}}{\partial n} 
= T^* \, \ln \left( \lambda_0(T^*) \frac{N-n}{L} \right)  \nonumber \\
&=& T^* \ln \left( \frac{\lambda_0(T^*)}{D} \, \tilde\rho_{free} \right) \;,
\end{eqnarray}
where $\tilde \rho_{free} = D(N-n)/L$ is the dimensionless density of
cars  in the free flow phase. The chemical potential of the cluster
phase can be easily calculated from Eqs.~(\ref{eq:ln}), (\ref{eq:lnw}),
and~(\ref{eq:Fder}). It yields
\begin{eqnarray}
\mu_{cl} &=& -T^* \left\{ \ln \left( \frac{D}{\lambda_0 \tilde b} \right)
- \ln \left[ 1 + \tilde \rho^2 \left( 1- \frac{n}{N} \right)^2 \right] \right\} 
\nonumber \\
&=& -T^* \left\{ \ln \left( \frac{D}{\lambda_0 \tilde b} \right)
-\ln \left[1+ \tilde \rho_{free}^2 \right] \right\} \;.
\end{eqnarray}
It is remarkable that, in distinction to the liquid--gas system, the
chemical potential of the cluster phase is not an internal property of
this phase, since it depends on the outer parameter -- the density of
the surrounding free--flow phase $\tilde \rho_{free}$. The physical
interpretation of this fact is that the traffic flow is a driven system,
which approaches a stationary  rather than equilibrium state in the
usual sense. However. as we have shown here, various properties of this
stationary state can be described by equilibrium thermodynamics. 

Free energy of the cluster phase can be calculated consistently
from~(\ref{F_traff}), (\ref{F0_tr}), and~(\ref{Fideal_tr})
according to $F=F_{cl}+F_{id}$. The result is
\begin{eqnarray}
&&F_{cl}(T^*,L,N,n) 
= T^* N \left\{ - \frac{n}{N} \left( 2 + \ln \left( \frac{D}{\lambda_0 \tilde b} \right) 
\right) \right. \nonumber \\ 
&&+ \left. \frac{2}{\tilde \rho} \left[ \arctan \tilde \rho  
-  \arctan \left( \tilde \rho \left[1- \frac{n}{N} \right] \right) \right]
 \right. \\
&&- \left. \left( 1- \frac{n}{N} \right) \ln \left( 1+ \tilde \rho^2
\left[ 1- \frac{n}{N} \right]^2 \right) + \ln \left( 1+ \tilde \rho^2 \right) 
\right\} \nonumber \;.
\end{eqnarray}
It is consistent with $\mu_{cl} = \partial F_{cl}/ \partial n$.

\section{Conclusions}

\begin{enumerate}
\item
In the current paper we have shown how thermodynamics can be applied to
such a many--particle system as traffic flow, based on a microscopic
(car--following) as well as a mesoscopic (stochastic cluster)
description, in analogy to equilibrium physical systems like
supersaturated vapour forming liquid droplets. The basic idea here is to
derive the free energy function and chemical potentials by using the
detailed balance, which holds in the stationary state of traffic flow in
the space of car cluster sizes.

\item
Distinguishing features between the traffic flow and equilibrium
physical systems have been outlined. In particular, we have found that
the third Newton's law does not hold on the level of "microscopic"
equations of motion for individual cars. Besides, the traffic flow is a
dissipative system with inflow and outflow of total energy. Unlike in
equilibrium systems, the chemical potential of the phase of congested
cars is not an internal property of this phase, since the traffic flow is
a driven system.
\end{enumerate}

\section*{Acknowledgments}
We would like to thank the German Science Foundation (DPG) for financial
support through grant MA~$1508/8-1$ and the Swedish Research Council 
through grant 2001-2545. We (J.~H.) also gratefully
acknowledge support by Graduiertenkolleg~567 \textit{Strongly Correlated
Many--Particle Systems} as well as (J.~K.) by Academic Exchange Program
(AAA) during the stay at Rostock University. The authors thank 
P.~Wagner (Berlin), St.~Trimper (Halle) and D.~Klochkov (Moscow)
for fruitful discussions.


\begin{thebibliography}{}

\bibitem{Grmela93}
	M. Grmela,
	Phys. Rev. E \textbf{48}, 919 (1993)

\bibitem{RHM86}
	H. Reiss, A. D. Hammerich, E. W. Montroll, 
	J. Stat. Phys. \textbf{42}, 647 (1986)

\bibitem{Smith05}
	E. Smith,
	Phys. Rev. E \textbf{72}, 036130 (2005)

\bibitem{LRM02}
	M. E. L\'arraga, J. A. del R\'io, A. Mehta, 
	Physica A \textbf{307}, 527 (2002)

\bibitem{vch99} 
	J. Schmelzer, G. R\"opke, R. Mahnke,
	\textit{Aggregation Phenomena in Complex Systems},
	(Wiley--VCH, Weinheim, 1999)

\bibitem{schw97}
	F. Schweitzer (ed.), \textit{Self--Organization of Complex Structures},
	(Gordon and Breach Science Publ., Amsterdam, 1977)

\bibitem{usms88} 
	H. Ulbricht, J. Schmelzer, R. Mahnke, F. Schweitzer, 
	\textit{Thermodynamics of Finite Systems and 
	the Kinetics of First--Order Phase Transitions},
	(Teubner, Leipzig, 1988)

\bibitem{prhe71}
	I. Prigogine, R. Herman, 
	\textit{Kinematic Theory of Vehicular Traffic}, 
	(Elsevier, New York, 1971)

\bibitem{helb04} 
	M. Krbalek, D. Helbing,
	Physica A \textbf{333}, 370 (2004)

\bibitem{helb06} 
	D. Helbing, M. Treiber, A. Kesting,
	Physica A \textbf{363}, 62 (2006)

\bibitem{makau99} R. Mahnke, J. Kaupu\v{z}s,
        Phys. Rev. E \textbf{59}, 117 (1999)

\bibitem{makaufr03} 
	R. Mahnke, J. Kaupu\v{z}s, V. Frishfelds,
	Atmospheric Research \textbf{65}, 261 (2003)

\bibitem{physrep} 
	R. Mahnke, J. Kaupu\v{z}s, I. Lubashevsky,
        Physics Reports \textbf{408}, Issue 1--2, pp. 1--130 (2005)

\bibitem{kue02} 
	R. K\"uhne, R. Mahnke, I. Lubashevsky, J. Kaupuzs, 
	Phys. Rev. E \textbf{65}, 066125, (2002).

\bibitem{Mukamel00}
	D.~Mukamel, Phase transitions in nonequilibrium systems, in 
	\textit{Soft and Fragile Matter: Nonequilibrium Dynamics,
	Metastability and Flow}, edited by M.~E. Cates and M.~R. Evans, 
	(Institute of Physics Publ., Bristol, 2000)

\bibitem{Schutz03}
	G.~M. Sch\"utz, J. Phys. A: Math. Gen. \textbf{36}, R339 (2003)

\bibitem{S1} 
	Y. Kafri, E. Levine, D. Mukamel, G. M. Sch\"utz, J. T\"or\"ok,
	Phys. Rev. Lett. \textbf{89}, 035702 (2002)

\bibitem{Evans05}
	M.~R. Evans, T.~Hanney,
	J. Phys. A: Math. Gen. \textbf{38}, R195 (2005)

\bibitem{KMH05}
	J. Kaupu\v{z}s, R. Mahnke, R. J. Harris,
	Phys. Rev. E \textbf{72}, 056125 (2005)

\bibitem{TGF98} 
	M. Schreckenberg, D. E. Wolf (Eds.), 
	\textit{Traffic and Granular Flow '97}, 
	(Springer, Singapore, 1998)

\bibitem{TGF96} 
	D. E. Wolf, M. Schreckenberg, A. Bachem (Eds.),
        \textit{Traffic and Granular Flow}, 
	(World Scientific Publ., Singapore, 1996)

\bibitem{cssrev} 
	D. Chowdhury, L. Santen and A. Schadschneider, 
	Physics Reports \textbf{329}, 199 (2000)

\bibitem{he01}
	D. Helbing, 
	Rev. Mod. Phys., \textbf{73}, 1067 (2001)

\bibitem{kerner04}
	B.~S. Kerner, \textit{The Physics of Traffic},
	(Springer, Berlin, 2004)

\bibitem{lumawa02} 
	I. Lubashevsky, R. Mahnke, P. Wagner, S. Kalenkov, 
	Phys. Rev. E \textbf{66}, 016117 (2002)

\bibitem{wagner2006}
	P. Wagner,
	European Phys. J. B \textbf{52}, 427 (2006) 

\bibitem{Arnold98} 
	L. Arnold, \textit{Random Dynamical Systems}, 
	(Springer, Berlin, 1998)

\bibitem{schi02} 
	V.~A. Anishenko, V.~V. Astakhov, A.~B. Neiman, 
	T.~E. Vadivasova, L. Schimansky-Geier, 
	\textit{Nonlinear Dynamics of Chaotic and Stochastic Systems},
	(Springer, Berlin, 2002)

\bibitem{erdeb00} 
	U. Erdmann, W. Ebeling, L. Schimansky-Geier, F. Schweitzer, 
	European Phys. J. B \textbf{15}, 105 (2000)

\bibitem{EbSo05} 
	W. Ebeling, I. M. Sokolov, 
	\textit{Statistical Thermodynamics and Stochastic Theory 
	of Nonequilibrium Systems}, 
	(World Scientific Publ., Singapore, 2005)

\bibitem{schw03} 
	F. Schweitzer, \textit{Brownian Agents and Active Particles}, 
	(Springer, Berlin, 2003)

\bibitem{betal94} 
	M. Bando, K. Hasebe, A. Nakayama, A. Shibata, Y. Sugiyama,
        Japan J. Indust. and Appl. Math. \textbf{11}, 203 (1994)

\bibitem{betal95a} 
	M. Bando, K. Hasebe, A. Nakayama, A. Shibata, Y. Sugiyama,
	Phys. Rev. E \textbf{51}, 1035 (1995)

\bibitem{betal95b} 
	M. Bando, K. Hasebe, K. Nakanishi, A. Nakayama, A. Shibata, Y. Sugiyama,
        J. Phys. I France \textbf{5}, 1389 (1995)

\bibitem{mapi97} R. Mahnke, N. Pieret,
        Phys. Rev. E \textbf{56}, 2666 (1997)

\bibitem{makau01} 
	R. Mahnke, J. Kaupu\v{z}s,
	Networks and Spatial Economics \textbf{1}, 103 (2001)

\end{thebibliography}
\end{document}